\documentclass{aa}
\usepackage{graphicx}
\usepackage{natbib}  
\usepackage{aalongtable}
\usepackage{txfonts}

\begin{document}
   \title{GIANO-TNG spectroscopy of red supergiants in the young star cluster RSGC2}

   \author{L. Origlia\inst{1}
          \and E. Oliva \inst{2}   
          \and R. Maiolino \inst{3} 
	  \and A. Mucciarelli \inst{4} 
	  \and C. Baffa \inst{2}
          \and V. Biliotti \inst{2}
          \and P. Bruno \inst{5} 
          \and G. Falcini  \inst{2}
          \and V. Gavriousev \inst{2}
          \and F. Ghinassi \inst{6} 
          \and E. Giani\inst{2}
          \and M. Gonzalez\inst{6}
          \and F. Leone\inst{7}   
          \and M. Lodi\inst{6}
          \and F. Massi\inst{2}
          \and P. Montegriffo\inst{1}
          \and I. Mochi\inst{8} 
          \and M. Pedani\inst{6}
          \and E. Rossetti\inst{4}
          \and S. Scuderi\inst{5}
          \and M. Sozzi\inst{2}
          \and A. Tozzi\inst{2}
          }

\institute{
             INAF - Osservatorio Astronomico di Bologna,
             Via Ranzani 1, I-40127 Bologna, Italy
	     \email{livia.origlia@oabo.inaf.it}
         \and
              INAF - Osservatorio Astrofisico di Arcetri,
              Largo E. Fermi 5, I-50125, Firenze, Italy
         \and
             University of Cambridge, Cavendish Lab.,
             JJ Thomson Av., Cambridge CB3 0HE, UK
         \and
             University of Bologna, Physics \& Astronomy Dept.,
             Viale Berti Pichat 6-2, I-40127 Bologna, Italy
         \and
             INAF - Osservatorio Astrofisico di Catania,
              via S. Sofia 78, I-95123 Catania, Italy
         \and
             INAF - TNG, ORM Astronomical Observatory,
              E-38787 Garafia, TF, Spain
         \and
             University of Catania, Physics \& Astronomy Dept.
              via S. Sofia 78, I-95123 Catania, Italy
         \and
             Lawrence Berkeley National Laboratory,
             1 Cyclotron Road, MS 2-400, Berkeley, CA 94720, USA
             }

\authorrunning{Origlia et al.}
\titlerunning{GIANO-TNG spectroscopy of red supergiants in RSGC2}

   \date{Received .... ; accepted ...}

 
  \abstract
   {}
   {The inner disk of the Galaxy has a number of young star clusters dominated by red supergiants that are 
heavily obscured by dust extinction and observable only at infrared wavelengths. 
These clusters are important tracers of the recent star formation and chemical enrichment history 
in the inner Galaxy.
   }
   {During the technical commissioning and as a first science verification 
of the GIANO spectrograph at the Telescopio Nazionale Galileo, 
we secured high-resolution (R$\simeq$50,000) near-infrared spectra 
of three red supergiants in the young Scutum cluster RSGC2. 
   }
   {Taking advantage of the full YJHK spectral coverage of GIANO in a single exposure, we were able to identify 
several tens of atomic and molecular lines suitable for chemical abundance determinations.
By means of spectral synthesis and line equivalent width measurements,  
we  obtained abundances of Fe and other iron-peak elements such as V, Cr, Ni,
of alpha (O, Mg, Si, Ca and Ti) and other light elements (C, N, Na, Al, K, Sc), and
of some s-process elements (Y, Sr).
We found iron abundances between half and one third solar and solar-scaled [X/Fe] abundance patterns 
of iron-peak, alpha and most of the  light elements,
consistent with a thin-disk chemistry.
We found a depletion of [C/Fe] and enhancement of [N/Fe], consistent with CN burning, and 
low $\rm ^{12}C/^{13}C$ abundance ratios (between 9 and 11), 
requiring extra-mixing processes in the stellar 
interiors during the post-main sequence evolution.   
Finally, we found a slight [Sr/Fe] enhancement and a slight [Y/Fe] depletion 
(by a factor of $\le$2), with respect to solar.   
   }
   {}


   \maketitle
%

\section{Introduction}

  \begin{figure*}
  \centering
  \includegraphics[width=\hsize]{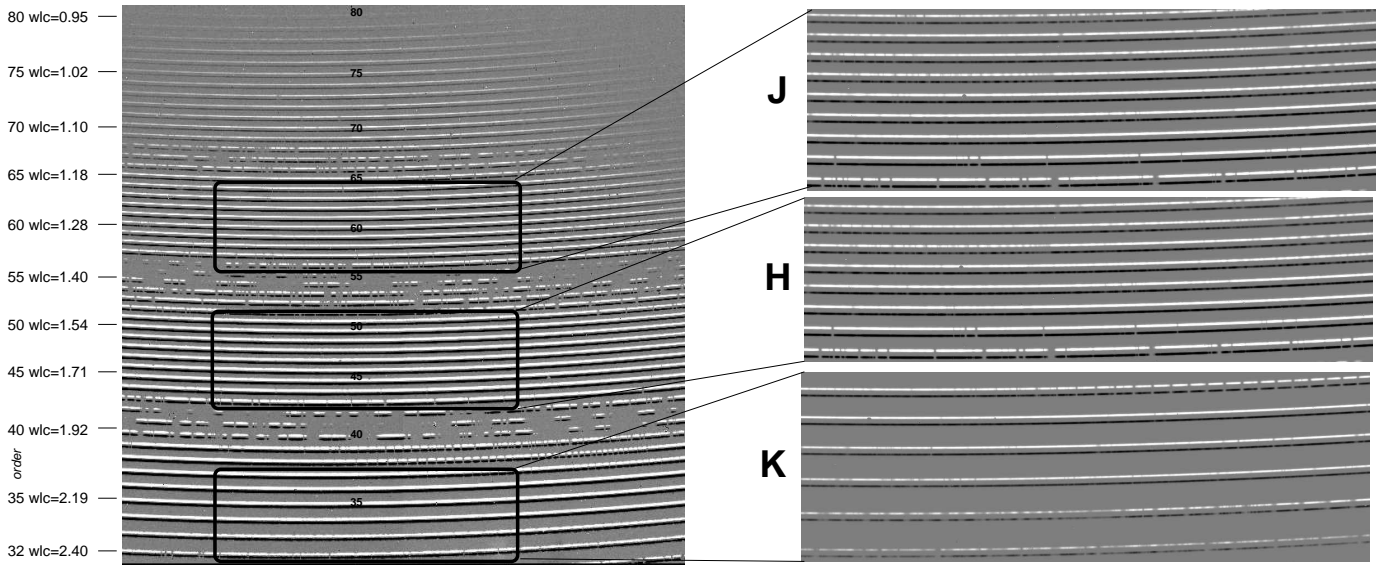}
   \caption{GIANO 2D (sky-subtracted) spectra of one of the observed RSG stars. Sky subtraction has been performed by 
nodding on fiber, resulting in one positive and one negative spectrum.} 
              \label{echelle}
    \end{figure*}

High-resolution spectroscopy in the near-infrared (NIR) is a 
powerful tool for measuring chemical abundances of cool giant and supergiant stars.
It is especially critical in very reddened environments such as the inner Galaxy, 
where extinction is so severe as to prevent any reliable measurement at shorter 
wavelengths.

Recently, our team made the first technical commissioning of GIANO, 
a cross-dispersed NIR spectrograph for the Telescopio Nazionale Galileo (TNG) 
at the Roque de Los Muchachos Observatory in La Palma (Spain). 
GIANO delivers a spectrum that covers
in a single exposure
the wavelength range  from 0.95~$\mu$m to 2.4~$\mu$m at a resolving power
R$\simeq$50,000.
The main disperser is a commercial R2 echelle grating
with 23.2 lines/mm working in quasi-Littrow configuration on
a $d$=100mm collimated beam. Cross dispersion is achieved via a network
of fused silica and ZnSe prisms that work in double pass, 
that is, they cross-disperse the light both before and after it is dispersed by
the echelle gratings, thus producing curved spectral orders.
The echellogram on the 2k$\times$2k detector spans 49 orders, from \#32 to \#80.
The spectral coverage is complete up to 1.7 $\mu$m.
At longer wavelengths the orders become larger than the
detector. The effective spectral coverage in the K-band is about 75\%.
Light from the telescope feeds a bundle of two IR-transmitting ZBLAN fibers,
with a core of 85~$\mu$m, corresponding to a sky-projected angle of 1~arcsec.  
The two fibers are aligned and mounted inside a custom connector.
The cores are at a distance of 0.25~mm, equivalent to a sky-projected angle of about 3~arcsec.

Owing to the constraints set by the visitor focus,
the fiber entrance was coupled to the TNG using a provisional, simplified
focal adapter that consisted of a commercial CaF$_2$ singlet lens positioned
26~mm before the fibers.
The focal adapter was mechanically mounted at a fiducial position, no
further adjustment of the optical axis was possible. This unfortunately
resulted in a very reduced efficiency of the system, and 
only bright targets were observable at that time.
More technical details on the instrument can be found in
\citet{oli12a,oli12b,oli13}.

As a first science verification of the GIANO performances, 
we observed three bright red supergiants (RSGs) in the star cluster RSGC2. 
RSGC2 is one of the few young, massive (4x10$^4$ M$_{\odot}$) clusters 
in the Galactic plane, located at the base of the Scutum-Crux arm 
and at the tip of the Galactic bar \citep{davies07} at a distance of $\simeq$3.5 kpc from the Galactic center.
Such rare clusters are rich in RSG stars and represent ideal laboratories for studying 
the evolution of massive stars as well as for 
physically and chemically characterizing 
the most recent and violent star formation events in the inner Galaxy. 
However, one can only study them in the IR or at radio wavelengths, because of high visual extinction 
($\rm A_V\approx 10-30$ mag) that affects the Galactic plane. 
Chemical abundances of Fe, C, and alpha-elements have been measured in 12 RSG stars, members of RSGC2,
by using medium-resolution (R$\simeq$17,000) spectra obtained with NIRSPEC at the KeckII telescope 
\citep[][hereafter D09]{davies09b}.
In some of these RSGs \citet{ver12} also found SiO maser emission.

We present and discuss the main atomic and molecular lines identified in the GIANO spectra of 
the observed three RSGs, which are members of the RSGC2 star cluster and are in common 
with the D09 sample. We also report the inferred chemical abundances of CNO and F from molecular lines, 
iron-peak, alpha, and other light elements and a few s-process elements from atomic lines.   

\section{Observations and data reduction}

For the observed targets, the 2MASS reference name, 
coordinates, spectral type and magnitudes, reddening and radial velocities from \citet{davies07} 
are reported in Table~\ref{tab1}. 

The reddening estimates do not affect the line equivalent width measurements since at NIR wavelengths 
the continuum is still dominated by the stellar photosphere and reddening dilutes 
both the line and the adjacent continuum flux
by the same amount.  

\begin{table*}
\caption{RSG stars observed with GIANO. Identification names, coordinates, and magnitudes are taken from 2MASS, 
reddening A$\rm _K$ and radial velocities are taken from \citet{davies07}.}
\label{tab1}     
\centering                          
\begin{tabular}{lllllllllll}  
\hline\hline
Ref & D09 Ref & ID & RA (J2000) & DEC (J2000) & SpT & J & H & K & A$\rm _K$ & RV (km/s)\\
\hline
\#1 & \#6 & J18391838-0600383 &  18~39~18.386 &-06~00~38.39 &  M5I & 7.717 & 5.919 & 5.072 & 1.17 & 107.1\\ 
\#2 & \#2 & J18391961-0600408 &  18~39~19.616 &-06~00~40.83 &  M3I & 6.899 & 5.045 & 4.120 & 1.39 & 111.1\\ 
\#3 & \#3 & J18392461-0602138 &  18~39~24.611 &-06~02~13.80 &  M4I & 7.273 & 5.458 & 4.499 & 1.34 & 110.5\\ 
\hline
\end{tabular}
\end{table*}

  \begin{figure}
  \centering
  \includegraphics[width=\hsize]{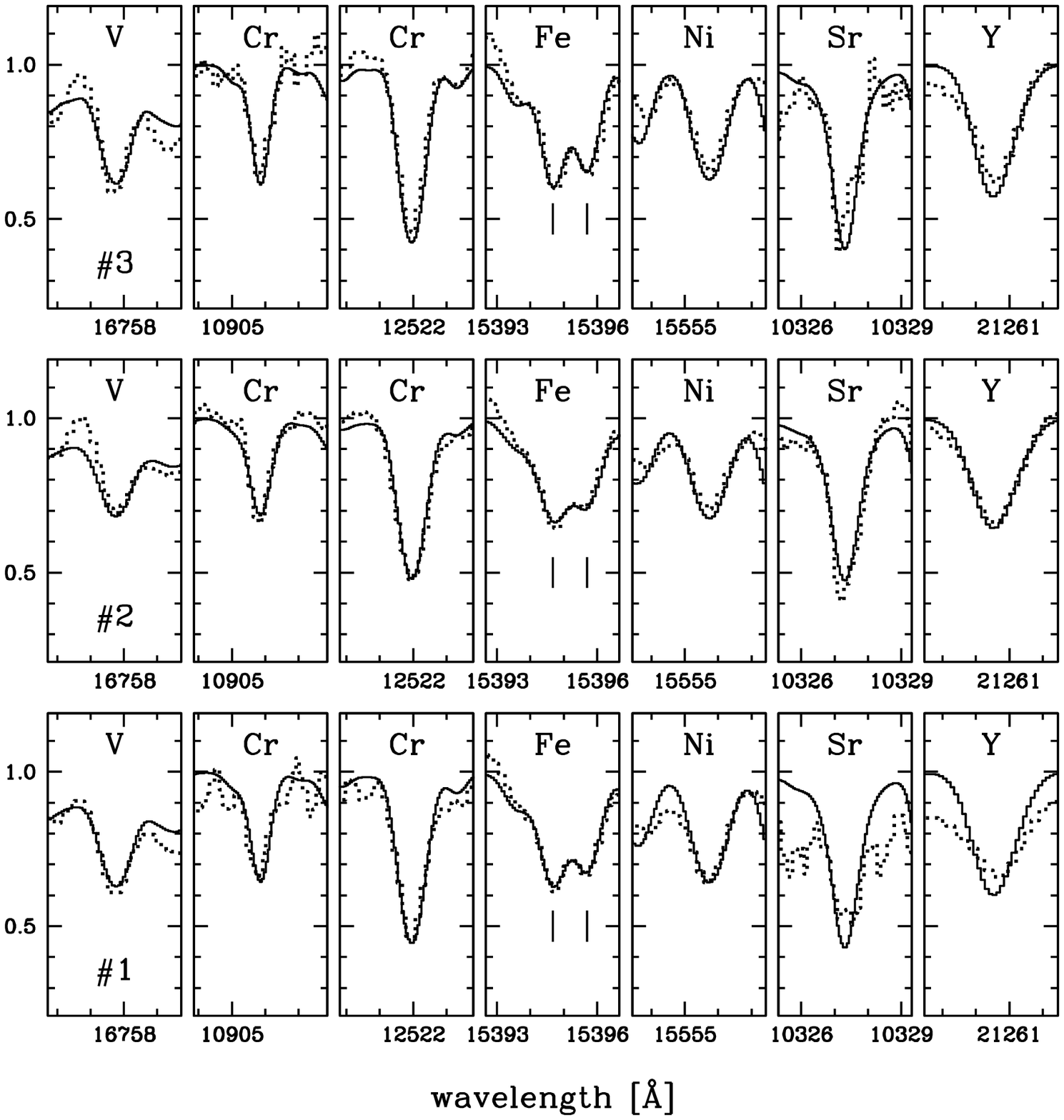}
   \caption{GIANO spectra around some iron-peak and s-process element lines
            for the three observed RSG stars (dotted lines). Our best-fit models are overplotted as solid lines.
            Rest frame wavelenghts are defined in air and thickmarks are every \AA\ in each panel.}
              \label{figiron}
    \end{figure}

  \begin{figure}
  \centering
  \includegraphics[width=\hsize]{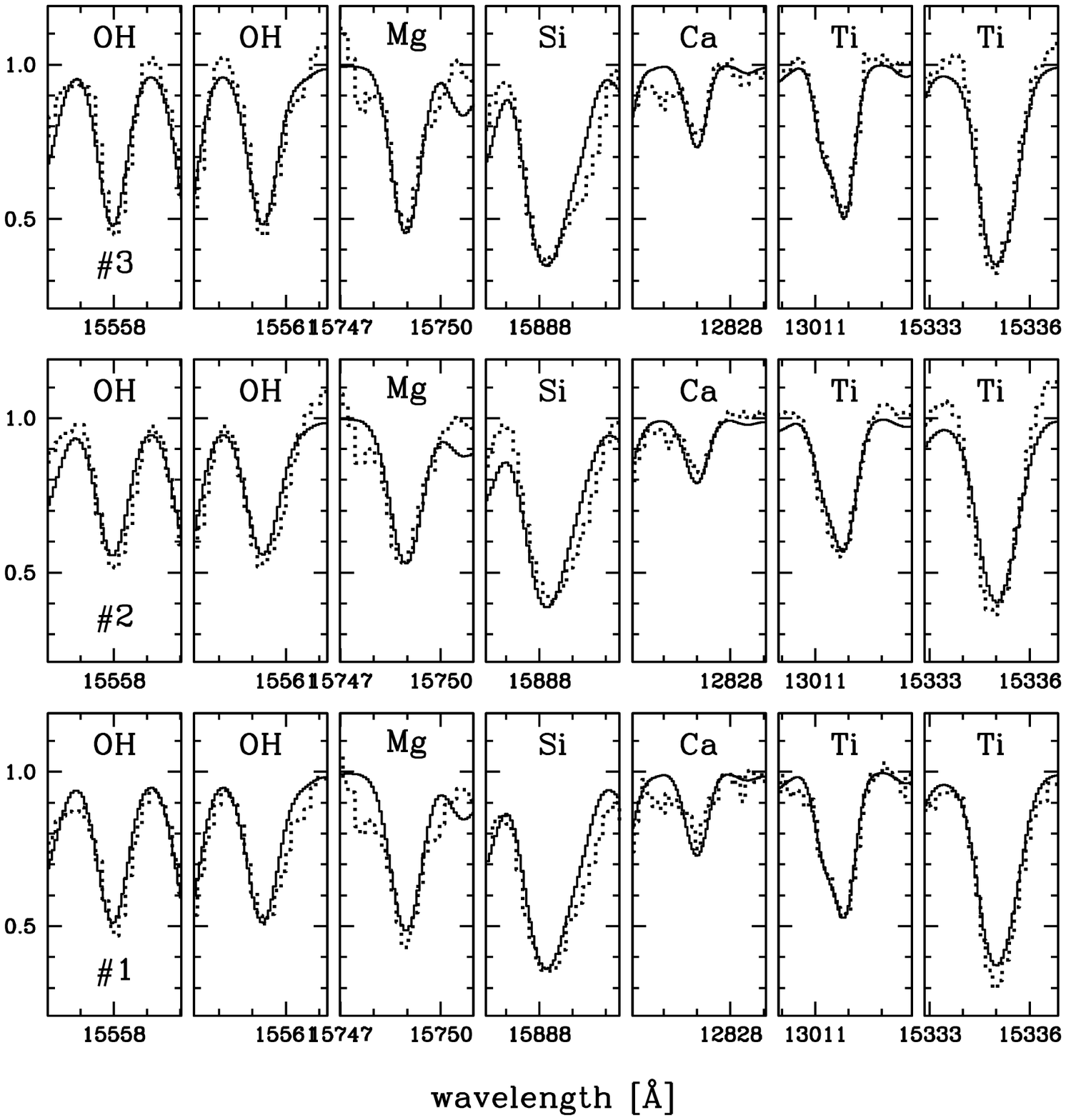}
   \caption{GIANO spectra around some alpha element lines
            for the three observed RSG stars (dotted lines). Our best-fit models are overplotted as solid lines.
	    Rest frame wavelengths are defined in air and thickmarks are every \AA\ in each panel.}
              \label{figalpha}
    \end{figure}

  \begin{figure}
  \centering
  \includegraphics[width=\hsize]{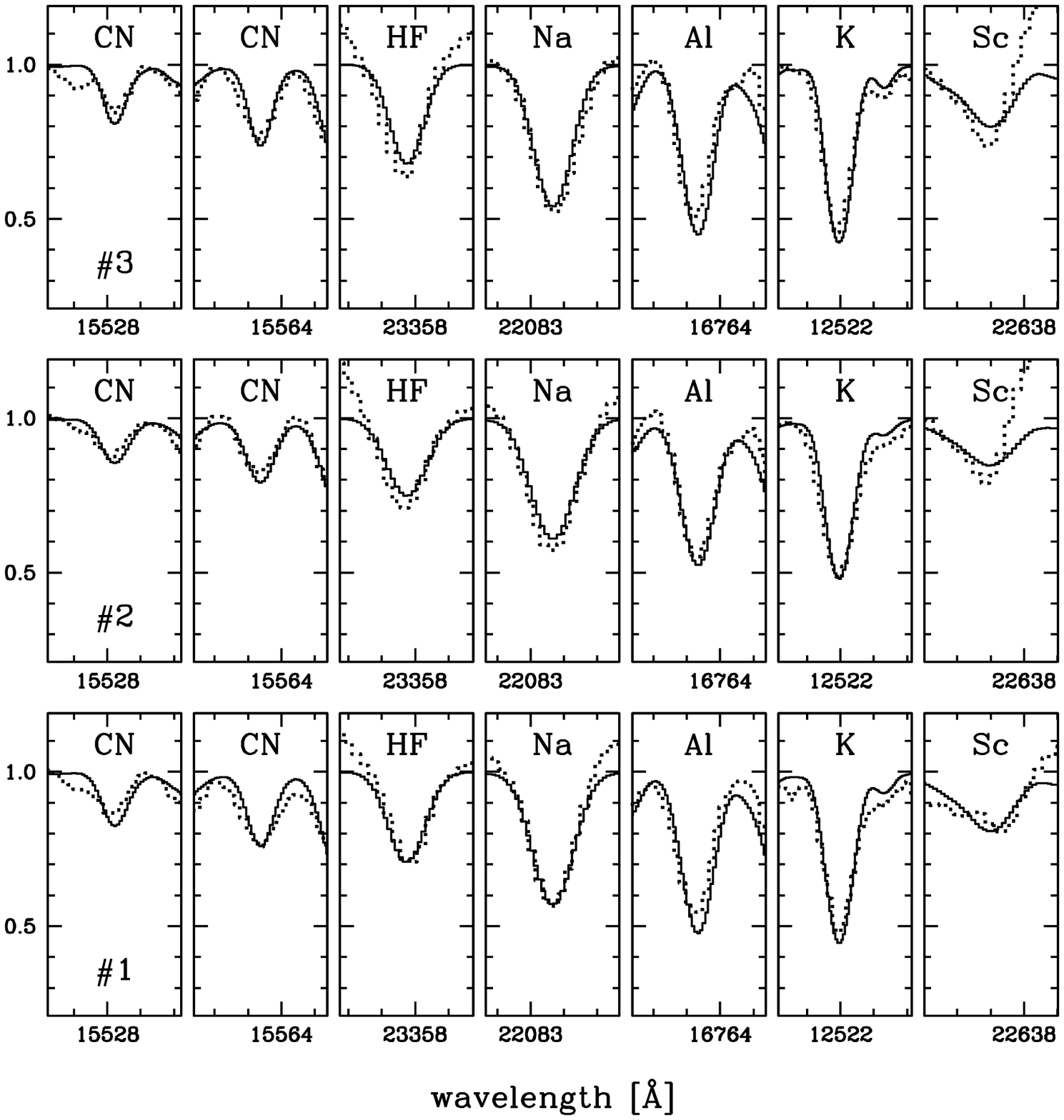}
   \caption{GIANO spectra around some light element lines
            for the three observed RSG stars (dotted lines). Our best-fit models are overplotted as solid lines.
	    Rest frame wavelengths are defined in air and thickmarks are every \AA\ in each panel.}
              \label{figlight}
    \end{figure}

The data were collected during the technical
night on July 30 2012. A first exposure of 5 min  
was acquired by centering the star+sky on fiber \#1 and the sky alone  on fiber \#2.
Then we acquired
a second exposure with the same integration time  but 
with star+sky on fiber \#2 and the sky alone on fiber \#1 
by nodding the telescope.
By subtracting the 2D spectrum of the second exposure from the 2D spectrum of the first exposure 
we obtained the pure star spectrum (see Fig.~\ref{echelle}). 
Nodding on fiber is very effective to properly subtract sky emission and instrumental background. 
The geometry of the orders was determined using flat exposures with a
tungsten calibration lamp. The 2D spectrum was thus rectified and the
spectra from each order  were extracted by summing six pixels around each fiber, in the
direction perpendicular to dispersion.
Wavelength calibration was determined by feeding the fibers with the light from
a U-Ne lamp. The wavelengths of the uranium lines were taken from 
\citet{red11},  while for neon we used the table available
on the NIST \citep{kra12}.
The $\lambda $ {\it versus} pixel relationship was obtained starting from a
physical model
of the instrument. This procedure is part of the pipeline that we are
developing for the instrument.
The resulting wavelength accuracy was about $\lambda/300,000$
r.m.s., that is, 0.05~\AA\  for lines in the H-band, while the overall signal-to-noise ratio 
is about 50.
For the three observed stars we found radial velocities consistent with the values reported 
by \citet{davies07} and listed in Table~\ref{tab1}.

Fig.~\ref{figiron} shows a few examples of the observed spectra for iron-peak and s-process elements, 
Fig.~\ref{figalpha} for alpha elements and Fig.~\ref{figlight} for other light elements.

\section{Chemical abundance analysis}

Deriving chemical abundances of RSG stars is an intrinsically difficult task for several reasons, 
namely the high level of molecular blending and blanketing in their spectra, the degeneracy 
in the determination of surface parameters, and the various broadening effects due to micro and macro turbulence.
The use of high-resolution spectroscopy in the NIR significantly mitigates the problem of the molecular 
blending and blanketing, which remains critical at optical wavelengths. 

To measure chemical abundances from the GIANO spectra, 
we used full spectral synthesis techniques 
and equivalent width measurements of selected lines, 
sufficiently isolated, free from significant blending and/or 
contamination by telluric absorption and without strong wings.
The presence of possible telluric absorption was carefully 
checked on an almost featureless O-star (Hip89584) spectrum.

We computed a large grid of suitable synthetic spectra
to model RSG stars by varying the stellar parameters and the
element abundances.
We used the same code as in D09 to facilitate the comparison.
This is an updated version \citep{ori02} of the program 
first described in \citet{ori93}.

The code was also successfully used to obtain abundances of 
bulge field \citep[][ and references therein]{rich12} 
and globular cluster \citep[][ and references therein]{ori08} giants,
young clusters dominated by RSGs \citep{lars06,lars08}, 
and to study the chemical abundances of RSGs in the Galactic center \citep{davies09a}.
The code uses the LTE approximation, is based
on the molecular blanketed model atmospheres of
\citet{jbk80} at temperatures $\le $4000~K
and on the ATLAS9 models for temperatures above 4000~K, and it
includes thousands of NIR atomic transitions 
from the Kurucz database\footnote{http://www.cfa.harvard.edu/amp/ampdata/kurucz23/sekur.html},
\citet{bie73}, and \citet{mel99}, while
molecular data are taken from our \citep[][ and subsequent updates]{ori93} 
and B. Plez (private communications) compilations.
The reference solar abundances are taken from \citet{gre98}.

Equivalent width measurements of the OH and CN molecular lines in the H-band, and of one HF 
line in the K-band were used to determine the oxygen,  
nitrogen and fluorine abundances. 

\begin{figure}
  \centering
  \includegraphics[width=\hsize]{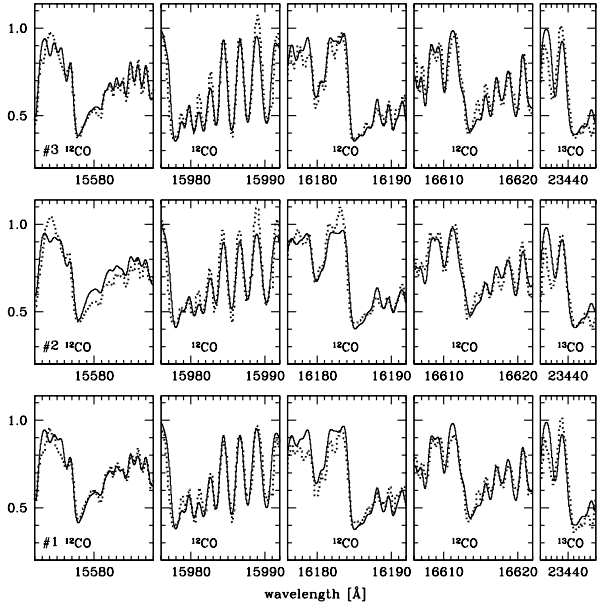}
   \caption{GIANO H-band spectra of the $^{12}$CO (3-0), (5-2), (6-3) and (8-5) bandheads and 
K-band spectra of the $^{13}$CO (3-0) bandhead  
            for the three observed RSG stars (dotted lines). Our best-fit models are overplotted as solid lines.
	    Rest frame wavelengths are defined in air.}
              \label{figco}
    \end{figure}

$^{12}$C and $^{13}$C carbon abundances were mostly determined from the CO bandheads in the 
H and K-bands, respectively, by means of full spectral synthesis, because of the high level of 
crowding and blending of the CO lines in these stars.
Fig.~\ref{figco} shows the GIANO spectra of the three RSGs centered on some of the CO bandheads used 
for determining the carbon abundances and our best-fit models.

Equivalent width measurements of neutral atomic lines over the full spectral range covered by the GIANO spectra 
were used to derive abundances of Fe and other iron-peak elements such as V, Cr, Ni,  
of alpha (Mg, Si, Ca and Ti) and other light elements such as Na, Al, K, Sc, and 
of the s-process element Y, while Sr abundances were obtained from ionized lines. 
Table~\ref{tab2} (available online) reports the list of lines used in the present analysis and their measured equivalent 
widths. 

Some other atomic lines of S (Y-band), Mn (J-band), Co, and Cu (H-band) are present in the spectra of the 
observed RSGs, but they are either too saturated or blended and/or contaminated by telluric absorption, 
which makes them ineffective for deriving reliable abundances.  

This compilation is a first census of the usable lines for abundance analysis and it does not pretend 
to be complete. 
Work is in progress to identify other potential lines of interest that, however, will be checked on 
forthcoming observed spectra of less extreme stars for a proper calibration. 
We note that our compilation contains several lines in common with the \citet{smi13} APOGEE 
spectral line list.

We compared the observed spectra with models in which temperature, gravity, and microturbulence 
velocity are as 
carefully determined in D09, that is T$_{\rm eff}$=3600~K, log~g=0.0 and $\xi$=2 km/s for all the three stars.
These parameters were fine-tuned to enable simultaneous spectral fitting of the CO and OH molecular bands
and the few atomic lines in the NIRSPEC spectra.
The many more CO, OH, and CN molecualr lines as well
as neutral atomic lines available in the GIANO spectra allowed us to 
carefully verify the reliability of the parameters adopted in D09.  
Models with these atmospheric parameters satisfactorily reproduce both the D09 and the current GIANO spectra. 
However, while D09 obtained good fits to the data without including an additional 
macroturbulence velocity broadening, since this was not resolved at their spectral resolution of R$\simeq$17,000,
this is not the case for the GIANO spectra at R$\simeq$50,000.
We obtained a good fit to the observed spectra by modeling macroturbulence velocity with 
a Gaussian $\sigma $ broadening of about 6.4, 7.1, and 5.7  km/s, or equivalently Doppler-broadening 
of 9, 10, and 8 km/s, for stars \#1, \#2 and \#3, respectively.
We did not find other appreciable line broadening by stellar rotation.

\begin{figure}
   \centering
  \includegraphics[width=\hsize]{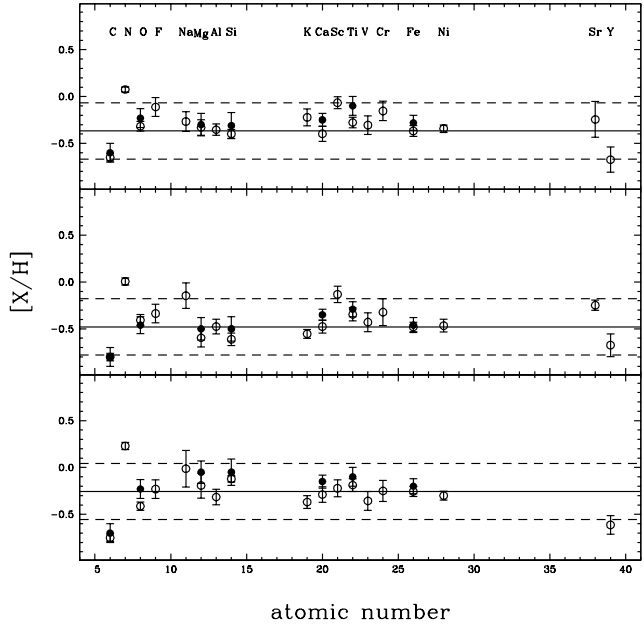}
   \caption{Open circles: [X/H] chemical abundances as a function of the atomic number for the three RSGs observed with GIANO. 
Solid dots: corresponding [X/H] abundances from D09. 
The horizontal lines mark the [Fe/H] abundance (solid line) and the $\pm$0.3 dex values (dashed lines),
as inferred in the present analysis.}
              \label{figabun}
    \end{figure}

The final average abundances are quoted in Table~\ref{tab3} 
and plotted in Fig.~\ref{figabun}.
The impact of using slightly different assumptions for the stellar parameters on the derived 
abundances is discussed in Sect.~\ref{error}.
 
Recently, \citet{dav13} discussed the problem of the most appropriate temperature scale for RSGs 
\citep[see also][]{lev05}, given that temperatures differing by several hundreds degrees can be inferred, 
using different scales. For the sake of consistency we therefore also explore models with significantly warmer ($\rm T_{eff}>$4000K) 
and cooler ($\rm T_{eff}<$3200K) temperatures than the one adopted in this analysis as best-fit value. 
We found that 
for temperatures above 3800K we were still iable to fit the observed spectra by using very peculiar
enhanced N and O abundances,
providing unlikely [O/Fe]$>$0.9 dex dex and [C+N/Fe]$>$1.0 dex. 
For temperatures below 3200K, we were barely (at $>$1.5 sigma level) able to fit the spectra with
very peculiar depleted N and O abundances, providing [O/Fe]$<-$0.3 dex 
and [C+N/Fe]$\approx$-0.5 dex. 
The impact on the overall iron and iron-peak elemental abundances is much weaker (within 0.1-0.2 dex).

\subsection{Error budget}
\label{error}

We can quantify random and systematic errors in the measurements of the equivalent widths and in the derived 
chemical abundances as follows:
the typical random error of the measured line equivalent widths is between 10 and 20 m\AA, mostly 
arising from a $\pm $2\% uncertainty in the placement of the pseudo-continuum, as estimated 
by overlapping the synthetic and the observed spectra. 
These random uncertainties in the line equivalent width measurements correspond to abundance variations 
ranging from a few hundredths to 1 tenth of a dex.
This $\le $0.1 dex error is lower than the typical 1$\sigma $ scatter in the derived abundances 
from different lines, which normally ranges between 0.1 and 0.2 dex.
The errors quoted in Table~\ref{tab3} for the final abundances were obtained 
by dividing these 1$\sigma $ errors by the squared root of the number of used 
lines, ranging from a few to about 20.    
A 10-20 m\AA\ variation of line equivalent width value was also obtained 
by varying the Gaussian line broadening (due to macro turbulence) by $\simeq$1 km/s, 
but this is mostly a systematic effect.

The other systematics arise from varying the adopted stellar parameters.
To properly quantify them, we generated a grid of test models with
T$_{\rm eff}$ between 3400~K and 3800~K, log~g between 0.0 and 1.0 dex,
and $\xi$ between 2 and 4 km/s.

We found that variations of $\pm $100~K with respect to the adopted temperature of 3600~K have 
a weak effect on the measured equivalent widths of atomic lines (on average the variation is $<$10 m\AA\, 
corresponding to a variation of few hundredths dex of the element abundance). 
Molecular OH and CN lines are more sensitive to temperature variations: their equivalent widths can vary by 15-20 m\AA\,
by varying temperature of $\pm $100~K,  corresponding to $\le $0.1 dex variation of the element abundance.    
A variation of log~g by $\pm $0.5 dex has a negligible impact on the equivalent width of the OH lines, while 
it produces a variation of 
$\simeq$20 m\AA\ ($\simeq$0.15 dex in abundance) and $\simeq$30 m\AA\ ($\simeq$0.20 dex in abundance) in the equivalent widths of the atomic and 
CN lines, respectively.
A variation of $\xi $ by $\pm $0.5 km/s has a negligible impact on the equivalent widths of CN lines, 
while it produces a variation of  
of $\simeq$15-20 m\AA\ in the equivalent widths of atomic and OH lines, 
corresponding to a variation of $\le $0.1 dex of the element abundance. 
A somewhat conservative estimate of the overall systematic uncertainty in the abundance (A) determination, caused by 
variations of the atmospheric parameters, can be computed as follows:
$\rm (\Delta A)^2 = (\partial A/\partial T)^2 (\Delta T)^2 + (\partial A/\partial log~g)^2 (\Delta log~g)^2  + (\partial A/\partial \xi)^2 (\Delta \xi)^2$
and it amounts to 0.15-0.20 dex. 

We also determined the statistical significance of our best-fit solution for the spectral synthesis of the 
CO features and the derived carbon abundances.
As a figure of merit of the statistical test we adopted
the difference between the model and the observed spectrum.
To quantify systematic discrepancies, this parameter is
more powerful than the classical $\chi ^2$ test, which is instead
equally sensitive to {\em random} and {\em systematic} scatters
\citep[see][ for more discussion and references]{ori04}.

Our best-fit solutions always show $>$90\% probability
to be representative of the observed spectra, while
those with $\pm$0.1~dex are significant at $\ge$1.5 $\sigma$ level only.

We also computed other test models with
$\rm \Delta T_{eff}=\pm$200~K, $\rm \Delta log~g=\pm$0.5~dex and
$\rm \Delta \xi=\pm$0.5~km~s$^{-1}$, and also with corresponding simultaneous variations
of the C abundance (0.1-0.2~dex) to reproduce the depth of the
molecular features.
CO molecular bands are very sensitive thermometers in the range
of temperature between 4500~K and 3800~K.
Indeed, temperature sets the fraction of molecular {\it versus} atomic
carbon.
At temperatures below ~3800~K most of the carbon 
is in molecular form, which drastically reduces the dependence of the CO 
band strengths on the temperature itself.
At temperatures $\ge$4500~K molecules barely survive, most of the
carbon is in atomic form, and the CO spectral features
become very weak.
Solutions with $\Delta log~g=\pm$0.5~dex or 
with $\Delta \xi=\pm$0.5~km/s 
and corresponding 0.1-0.2 dex variation of the carbon abundance 
are significant at $\ge$1.5 $\sigma$ level only. 

\section{Results and discussion}

\begin{table}
\caption{Chemical abundances of the RSG stars observed with GIANO.}
\label{tab3}     
\centering                          
\begin{tabular}{llllll}  
\hline\hline
Element & Atomic \# & \multicolumn{3}{c}{[X/H]} & n$_{\rm lines}$ \\ 
        &           & (\#1) & (\#2) & (\#3) & \\
\hline
Fe& 26 & -0.26 & -0.48 & -0.37 &  13 \\
 &&  $\pm$0.05 & $\pm$0.05 & $\pm$0.06 &  \\
V & 23 & -0.36 & -0.43 & -0.30 &   1 \\
 &&  $\pm$0.10 &  $\pm$0.10 &  $\pm$0.10 &  \\
Cr& 24 & -0.25 & -0.32 & -0.15 &   4 \\
 &&  $\pm$0.11 &  $\pm$0.14 &  $\pm$0.10 &  \\
Ni& 28 & -0.30 & -0.47 & -0.34 &   6 \\
 &&  $\pm$0.05 & $\pm$0.07 & $\pm$0.04 &  \\
Mg& 12 & -0.19 & -0.60 & -0.33 &   5 \\
 &&  $\pm$0.13 & $\pm$0.10 & $\pm$0.08 &  \\
Si& 14 & -0.12 & -0.61 & -0.40 &   7 \\
 && $\pm$0.03 & $\pm$0.07 & $\pm$0.05 &  \\
Ca& 20 & -0.29 & -0.48 & -0.40 &   7 \\
 && $\pm$0.08 & $\pm$0.07 & $\pm$0.08 &  \\
Ti& 22 & -0.19 & -0.35 & -0.28 &  21 \\
 && $\pm$0.07 & $\pm$0.07 & $\pm$0.06 &  \\
C &  6 & -0.75 & -0.85 & -0.65 &  - \\
 && $\pm$0.05 & $\pm$0.05 & $\pm$0.05 &  \\
N &  7 & +0.23 & +0.01 & +0.07 &  18 \\
 && $\pm$0.04 & $\pm$0.04 & $\pm$0.03 &  \\
O &  8 & -0.41 & -0.41 & -0.32 &  23 \\
 && $\pm$0.04 & $\pm$0.06 & $\pm$0.05 &  \\
F &  9 & -0.23 & -0.34 & -0.11 &   1 \\
 && $\pm$0.10 & $\pm$0.10 & $\pm$0.10 &  \\
Na& 11 & -0.01 & -0.15 & -0.27 &   2 \\
 && $\pm$0.20 & $\pm$0.14 & $\pm$0.11 &  \\
Al& 13 & -0.32 & -0.48 & -0.35 &   3 \\
 && $\pm$0.08 & $\pm$0.08 & $\pm$0.06 &  \\
K & 19 & -0.37 & -0.55 & -0.22 &   3 \\
 && $\pm$0.07 & $\pm$0.05 & $\pm$0.09 &  \\
Sc& 21 & -0.22 & -0.13 & -0.07 &   2 \\
 && $\pm$0.09 & $\pm$0.09 & $\pm$0.06 &  \\
Sr& 38 & - & -0.25 & -0.24 &   2 \\
 && - & $\pm$0.06 & $\pm$0.19 &  \\
Y & 39 & -0.61 & -0.67 & -0.67 &   2 \\
 && $\pm$0.10 & $\pm$0.12 & $\pm$0.13 &  \\
\hline\hline
$\rm ^{12}C/^{13}C$ & -& 9 & 10 & 11 & - \\
 && $\pm$1 & $\pm$1  & $\pm$1  &  \\
\hline
\end{tabular}
\end{table}

The abundances of all chemical elements with measurable lines in the GIANO spectra 
are very similar in all the three RSGs, as expected because they are members of a star cluster. 

We found iron abundances between half and one third solar, in good agreement with the previous estimates by D09. 
The other iron-peak elements (V, Cr, Ni) show abundances that are consistent with the iron values to within $\pm$0.15 dex,  
thus fully confirming the sub-solar metallicity of the cluster.   

The alpha (O, Mg, Si, Ca, Ti) as well as most of the other light (F, Na, Al,K, Sc) elements behave in a similar fashion, 
with [X/Fe] abundance ratios about solar ($\pm$0.15 dex); the only exceptions are [F/Fe] and [Sc/Fe] in star \#1, 
[Na/Fe] in star \#2 and 
[Na/Fe] and [Sc/Fe] in star \#3, which are slightly enhanced (but always within a factor of two).

Interestingly, the measured [F/O] abundance ratios  
(+0.18 in star \#1, +0.07 in star \#2 and +0.21 in star \#3), 
are fully consistent with the values measured in K-M dwarfs in the Orion Nebula Cluster 
by \citet{cun05} 
and are representative of the disk composition.

The Sr and Y s-process elements behave somewhat differently from each other: 
[Sr/Fe] is slightly enhanced, while [Y/Fe] is slightly depleted 
(by a factor of $\le$2) with respect to solar.   

[C/Fe] is depleted by a factor between two and three, similarly to what was found by D09, and the 
$\rm ^{12}C/^{13}C$ ratio is also rather low (between 9 and 11), 
indicating that some extra-mixing processes in the stellar 
interiors are at work during the post-main sequence evolution.   
Evolutionary tracks of massive stars with rotation \citep[e.g.][]{mey00} can account for both the overall 
depletion of carbon and the low $\rm ^{12}C/^{13}C$ isotopic ratio, and indeed, 
deep, rotationally enhanced mixing was previously suggested by D09.

[N/Fe] is enhanced in such a way that [C+N/Fe] is about 0.0 (+0.00 in star \#1, +0.12 in star \#2, and +0.16 in star \#3), 
consistent with standard CN nucleosynthesis. 

The mostly solar-scaled [X/Fe] abundance patterns of iron-peak, alpha, and other light elements 
as measured in the observed three RSGs of RSCG2 
are fully consistent with the chemistry of the thin disk \citep[see e.g.][]{reddy03}, 
that underwent chemical enrichment over long timescales with the contribution of both type II and type I supernovae.
At variance, the sub-solar metallicity of RSGC2 and of its companion cluster RSGC1 (see D09) is intriguing.
Indeed, abundance measurements 
of Cepheids in the inner disk 
\citep[see e.g.][ and references therein]{gen13,and13}
indicate the existence of a positive metallicity gradient toward the inner regions and  
metal abundances well in excess of solar. 
However, these measurements still sample a region at a minimum Galactocentric distance of about 4 kpc
because of the huge extinction closer to the center. 
Only a few IR measurements of RSGs exist so far
in the innermost disk region and in the Galactic center 
\citep[e.g.][]{ram00,mar08,naj09,davies09a,davies09b},
consistent with a metal abundance of about solar and some level of alpha-element enhancement 
\citep[see e.g.][]{cun07}.

It is therefore clear that neither  RSGs in the Scutum star clusters (D09 and the present work) nor RSGs  
in the center of the Galaxy
follow the radial Galactic trend traced by Cepheids and other young stellar populations at 
Galactocentric distances R${\rm GC}>$3-4~kpc \citep[see D09 and][ for more discussion and references]{gen13}. 
However, this is not surprising, because 
the innermost region of the Galaxy shows several different substructures, 
such as the bulge/bar and  the ends of the spiral arms, 
with strong fluctuations in stellar/gas density and star formation rates.
In such a complex physical and kinematic environment, the chemical enrichment 
process is also expected to have been quite inhomogeneous. 

\section{Conclusions}

As a preliminary test of the scientific performances of the GIANO spectrograph during its first 
commissioning in July 2012,
we observed three bright RSGs in the Scutum star cluster RSGC2.

A NIR spectrograph such as GIANO, which provides high spectral resolution and full spectral coverage 
in a single exposure, is a unique instrument to perform detailed chemical studies of cool stars  
in any Galactic environment, including highly reddened star clusters such as RSGC2 in the inner Galactic disk. 

The high spectral resolution has allowed us to resolve and measure the macro turbulence broadening 
in the observed RSGs and to better identify unblended lines for accurate equivalent width measurements and 
chemical abundance determinations.
 
The simultaneous access to the YJH and K bands offers the possibility 
of sampling most of the elements of interest for a complete check of the 
chemical clock and of measuring from a few to several tens of lines per element, thus enabling an accurate 
and statistically significant abundance analysis.  
We used the Y-band to measure Cr and Sr, the J-band to measure K and some Fe, Cr, Ca, Mg, Si, and Ti lines, 
the H-band to measure $^{12}$C, N, O, Al, V, and Ni, most of the Fe lines and some Al Ca, Mg, Si and Ti lines, 
and finally the K-band to measure $^{13}$C , F, Na, Sc, Y and a few Al, Ti and Fe lines.

For the three observed RSGs we found overall abundance patterns consistent with the thin-disk chemistry, and we confirmed 
the sub-solar metallicity of the RSGC2 star cluster, as previously suggested by D09, 
which indicates that the inner disk in particular, 
but more generally the inner Galaxy, has quite an inhomogeneous chemical composition.

\begin{acknowledgements}
Part of this work was supported by the grant TECNO-INAF-2011.\\
This publication makes use of data products from the Two Micron All Sky Survey, 
which is a joint project of the University of Massachusetts and the Infrared 
Processing and Analysis Center/California Institute of Technology, 
funded by the National Aeronautics and Space Administration and the National Science Foundation. 
We thanks Ben Davies for his carefull referee report.
\end{acknowledgements}

\Online
\begin{longtable}{llllll}  
\caption{Lines identified in the RSG stars observed with GIANO. Wavelength are defined in air 
($\rm \lambda_{air}=\lambda_{vacuum}/1.000274$). 
We also list the line transition probability (log gf), excitation potential ($\chi $) and the measured equivalent widths in the three 
observed RSG stars. \label{tab2}}\\
\hline\hline
$\rm \lambda _{air}$ & log gf & $\chi $ & EW(\#1) & EW(\#2) & EW(\#3)\\
$\mu $m    &        & eV      & m\AA    & m\AA    & m\AA \\
\hline
\endfirsthead
\caption{continued.}\\ 
\hline\hline
$\rm \lambda _{air}$ & log gf & $\chi $ & EW(\#1) & EW(\#2) & EW(\#3)\\
$\mu $m    &         & eV     & m\AA & m\AA & m\AA \\
\hline
\endhead
\multicolumn{6}{l}{Fe I lines}\\
1.22133 & -1.94 & 4.64 & 133 &  88 &  93 \\ 
1.22271 & -1.46 & 4.61 & 160 & 140 & 146 \\ 
1.23429 & -1.56 & 4.64 & 186 & 151 & 139 \\ 
1.26159 & -1.61 & 4.64 & 131 & 127 & 128 \\ 
1.28249 & -4.19 & 3.02 & 193 & 170 & 151 \\ 
1.28406 & -1.47 & 4.96 & 134 & 140 & 152 \\ 
1.52075 &  0.23 & 5.39 & 378 & 360 & 336 \\ 
1.52450 & -0.28 & 5.59 & 349 & 302 & 303 \\ 
1.53947 & -0.12 & 5.62 & 360 & 370 & 333 \\ 
1.53957 & -0.26 & 5.62 & 296 & 284 & 291 \\ 
1.57694 &  0.60 & 5.54 & 469 & 466 & 403 \\ 
1.59649 & -0.08 & 5.92 & 268 & 269 & 263 \\ 
2.23808 & -0.46 & 5.04 & 447 & 424 & 410 \\ 
\hline
\multicolumn{6}{l}{V I lines} \\
1.67577 & -1.37 & 3.41 & 296 & 291 & 288 \\ 
\hline
\multicolumn{6}{l}{Cr I lines} \\
1.06721 & -1.37 & 3.02 & 212 & 189 & 256 \\ 
1.08014 & -1.72 & 3.01 & 188 & 195 & 182 \\ 
1.09059 & -0.65 & 3.44 & 217 & 248 & 215 \\ 
1.25218 & -1.59 & 2.71 & 379 & 383 & 327 \\ 
\hline
\multicolumn{6}{l}{Ni I lines} \\
1.55554 &  0.07 & 5.49 & 320 & 272 & 271 \\ 
1.63105 &  0.07 & 5.29 & 333 & 338 & 299 \\ 
1.63642 &  0.44 & 5.29 & 474 & 482 & 454 \\ 
1.65893 & -0.49 & 5.47& 156 & 159 & 161 \\ 
1.69967 &  0.31 & 5.31 & 339 & 337 & 295 \\ 
1.70016 &  0.23 & 5.49 & 250 & 217 & 219 \\ 
\hline
\multicolumn{6}{l}{Mg I lines} \\
1.18282 & -0.29 & 4.35 & 392 & 402 & 377 \\ 
1.57407 & -0.24 & 5.94 & 419 & 382 & 374 \\ 
1.57490 & -0.06 & 5.94 & 486 & 452 & 405 \\ 
1.57658 &  0.38 & 5.94 & 490 & 482 & 434 \\ 
1.59545 & -1.03 & 6.59 & 287 & 239 & 249 \\ 
\hline
\multicolumn{6}{l}{Si I lines} \\
1.19916 & -0.16 & 4.92 & 308 & 258 & 241 \\ 
1.21035 & -0.39 & 4.93 & 283 & 278 & 247 \\ 
1.22707 & -0.41 & 4.96 & 279 & 248 & 240 \\ 
1.58884 & -0.03 & 5.09 & 554 & 567 & 493 \\ 
1.59601 &  0.13 & 5.99 & 417 & 386 & 370 \\ 
1.60600 & -0.44 & 5.96 & 311 & 239 & 285 \\ 
1.60948 & -0.11 & 5.97 & 371 & 315 & 324 \\ 
\hline
\multicolumn{6}{l}{Ca I lines} \\
1.28160 & -0.63 & 3.91 & 220 & 222 & 199 \\ 
1.28239 & -0.85 & 3.91 & 209 & 221 & 219 \\ 
1.28271 & -1.33 & 3.91 & 157 & 124 & 117 \\ 
1.61508 &  0.36 & 5.30 & 364 & 354 & 331 \\ 
1.61552 & -0.02 & 5.30 & 277 & 211 & 218 \\ 
1.61574 &  0.49 & 5.32 & 327 & 287 & 289 \\ 
1.61971 &  0.64 & 5.30 & 439 & 406 & 363 \\ 
\hline
\multicolumn{6}{l}{Ti I lines} \\
1.17805 & -2.38 & 1.44 & 296 & 241 & 224 \\ 
1.17972 & -2.46 & 1.43 & 289 & 309 & 273 \\ 
1.18929 & -1.91 & 1.43 & 326 & 356 & 305 \\ 
1.19495 & -1.76 & 1.44 & 303 & 328 & 279 \\ 
1.23884 & -2.20 & 2.16 & 166 & 134 & 135 \\ 
1.25696 & -2.26 & 2.18 & 205 & 167 & 208 \\ 
1.26711 & -2.52 & 1.43 & 244 & 322 & 304 \\ 
1.27384 & -1.39 & 2.18 & 258 & 283 & 262 \\ 
1.27449 & -1.35 & 2.49 & 259 & 220 & 190 \\ 
1.28217 & -1.53 & 1.46 & 411 & 439 & 360 \\ 
1.28314 & -1.11 & 1.43 & 382 & 381 & 349 \\ 
1.28470 & -1.55 & 1.44 & 382 & 401 & 355 \\ 
1.29199 & -1.22 & 2.15 & 258 & 209 & 226 \\ 
1.29876 & -1.78 & 2.51 & 143 & 167 & 161 \\ 
1.30054 & -2.46 & 2.18 & 161 & 155 & 173 \\ 
1.30119 & -2.49 & 1.44 & 313 & 317 & 280 \\ 
1.53348 & -1.15 & 1.89 & 596 & 608 & 512 \\ 
1.55438 & -1.48 & 1.88 & 558 & 580 & 497 \\ 
2.17829 & -1.16 & 1.75 & 735 & 726 & 648 \\ 
2.18974 & -1.45 & 1.74 & 644 & 615 & 595 \\ 
2.24439 & -2.25 & 1.74 & 536 & 478 & 486 \\ 
\hline
\multicolumn{6}{l}{Al I lines} \\
1.67190 &  0.15 & 4.09 & 477 & 478 & 446 \\ 
1.67634 & -0.55 & 4.09 & 421 & 423 & 394 \\ 
2.10930 & -0.31 & 4.09 & 509 & 503 & 475 \\ 
\hline
\multicolumn{6}{l}{Na I lines} \\
2.20837 & -0.02 & 3.19 & 538 & 548 & 494 \\ 
2.33791 &  0.53 & 3.76 & 566 & 563 & 489 \\ 
\hline
\multicolumn{6}{l}{K I lines} \\
1.17728 &  0.51 & 1.62 & 293 & 287 & 292 \\ 
1.24323 & -0.44 & 1.61 & 222 & 234 & 244 \\ 
1.25221 & -0.14 & 1.62 & 358 & 353 & 318 \\ 
\hline
\multicolumn{6}{l}{Sc I lines} \\
2.17305 & -1.68 & 1.44 & 447 & 513 & 490 \\ 
2.26370 & -2.17 & 1.44 & 313 & 367 & 337 \\ 
\hline
\multicolumn{6}{l}{Sr II lines} \\
1.03273 & -0.35 & 1.84 &  -  & 387 & 325 \\ 
1.09149 & -0.64 & 1.81 &  -  & 383 & 344 \\ 
\hline
\multicolumn{6}{l}{Y I lines} \\
2.12604 & -0.10 & 1.43 & 419 & 446 & 417 \\ 
2.25438 & -0.20 & 1.40 &  -  & 362 & 346 \\ 
\hline
\multicolumn{6}{l}{$\rm ^{12}C^{14}N$ lines} \\
1.52132 & -2.00 & 0.83 & 198 & 194 & 197 \\ 
1.52195 & -1.80 & 0.73 & 313 & 287 & 267 \\ 
1.52604 & -1.35 & 0.92 & 235 & 195 & 180 \\ 
1.52689 & -1.41 & 1.29 & 161 & 123 & 121 \\ 
1.52726 & -1.57 & 0.77 & 202 & 174 & 162 \\ 
1.52916 & -1.92 & 0.85 & 144 & 162 & 135 \\ 
1.53071 & -1.90 & 0.85 & 168 & 166 & 162 \\ 
1.53214 & -1.75 & 0.79 & 203 & 183 & 168 \\ 
1.53974 & -1.80 & 0.91 & 227 & 211 & 177 \\ 
1.54396 & -1.77 & 0.93 & 184 & 193 & 176 \\ 
1.54471 & -1.16 & 1.09 & 217 & 214 & 196 \\ 
1.54662 & -1.15 & 1.09 & 238 & 230 & 205 \\ 
1.55282 & -1.71 & 1.00 & 143 & 117 & 126 \\ 
1.55634 & -1.14 & 1.15 & 219 & 139 & 193 \\ 
1.56093 & -1.51 & 0.94 & 217 & 199 & 203 \\ 
1.56136 & -1.14 & 1.18 & 282 & 225 & 224 \\ 
1.56736 & -1.65 & 1.07 & 288 & 207 & 221 \\ 
1.56749 & -1.65 & 1.09 & 208 & 164 & 198 \\ 
\hline
\multicolumn{6}{l}{$\rm ^{16}O^{1}H$ lines} \\
1.51458 & -5.63 & 0.16 & 425 & 478 & 427 \\ 
1.51479 & -5.63 & 0.16 & 391 & 461 & 400 \\ 
1.52366 & -5.93 & 0.45 & 260 & 349 & 238 \\ 
1.53285 & -5.67 & 0.47 & 451 & 441 & 378 \\ 
1.53912 & -5.59 & 0.49 & 330 & 354 & 325 \\ 
1.54092 & -5.55 & 0.26 & 376 & 451 & 401 \\ 
1.54698 & -5.26 & 0.94 & 220 & 250 & 309 \\ 
1.54969 & -5.26 & 0.91 & 302 & 306 & 267 \\ 
1.55053 & -5.46 & 0.52 & 358 & 438 & 383 \\ 
1.55580 & -5.49 & 0.30 & 452 & 461 & 417 \\ 
1.55602 & -5.49 & 0.30 & 422 & 455 & 418 \\ 
1.55658 & -5.42 & 0.90 & 234 & 227 & 228 \\ 
1.55688 & -5.45 & 0.30 & 395 & 432 & 409 \\ 
1.55721 & -5.45 & 0.30 & 418 & 463 & 418 \\ 
1.56519 & -5.29 & 0.53 & 324 & 270 & 273 \\ 
1.56535 & -5.29 & 0.53 & 389 & 366 & 345 \\ 
1.58977 & -5.36 & 0.41 & 442 & 465 & 434 \\ 
1.59104 & -5.14 & 0.60 & 359 & 338 & 357 \\ 
1.59127 & -5.14 & 0.60 & 370 & 373 & 363 \\ 
1.63522 & -5.00 & 0.74 & 448 & 442 & 442 \\ 
1.63546 & -5.00 & 0.74 & 365 & 333 & 353 \\ 
1.63681 & -4.96 & 0.73 & 359 & 331 & 368 \\ 
1.68723 & -5.15 & 0.76 & 372 & 353 & 388 \\ 
\hline
\multicolumn{6}{l}{$\rm ^{1}H^{9}F$ lines} \\
2.33577 & -3.95 & 0.48 & 372 & 353 & 388  \\
\hline
\end{longtable}
\end{document}